\documentclass[9pt,twocolumn,twoside]{opticajnl}
\journal{opticajournal} 

\setboolean{shortarticle}{false}

\usepackage{commath}

\usepackage{lineno} 

\usepackage{siunitx}
\usepackage{comment}
\newcommand{\SIadj}[2]{\SI[number-unit-product={\text{-}}]{#1}{#2}}
\DeclareSIUnit\bar{bar}

\newcommand{\ie}{\textit{i}.\textit{e}.,\ }
\newcommand{\eg}{\textit{e}.\textit{g}.,\ }
\title{Raman-enhanced spectral compression of high-energy femtosecond laser pulses in molecular gases}

\newcommand{\RN}[1]{%
\textup{\uppercase\expandafter{\romannumeral#1}}%
} 
\newcommand{\rn}[1]{%
\textup{\lowercase\expandafter{\romannumeral#1}}%
} 

\usepackage[version=4]{mhchem} 

\author[1,$\dagger$]{Zegui Wang}
\author[2,$\dagger$]{Yi-Hao Chen}
\author[3]{Yunlong Mo}
\author[3]{Zaitian Dong}
\author[3]{Wanhong Yin}
\author[2]{Frank Wise}
\author[1,*]{Wei Cao}

\affil[1]{School of Physics and Wuhan National Laboratory for Optoelectronics, Huazhong University of Science and Technology, Wuhan 430074, China}
\affil[2]{School of Applied and Engineering Physics, Cornell University, Ithaca, New York 14853, USA}
\affil[3]{Xi'an Institute of Applied Optics, Xi'an 710065, China}
\affil[$\dagger$]{These authors contributed equally to this work.}

\affil[*]{weicao@hust.edu.cn}

\begin{abstract}
Nonlinear pulse propagation in gas-filled waveguides has attracted substantial attention over the past decade, and a variety of capabilities have been reported. However, there is no prior report of spectral compression in gas-filled waveguides or cavities, which would offer a natural route for scaling to much higher pulse energies than have been reached in solid structures. Here we report a high-energy spectral-compression technique based on nonlinear propagation in gas-filled capillaries. With \num{0.1}- to \SIadj{1}{\milli\joule} pulses, compression of the spectral width by a factor up to \num{12} (from \SI{60}{\nm} to \SI{5}{\nm}) is demonstrated. Key to this advance is recognition that the process plays out differently in gases than in solids. In a noble gas (\ce{Ar}), we find that even small structure in the spectrum, which is mapped to the time profile, of the input pulse can degrade the compression process. We identify the delayed Raman response of molecular gases (\ce{N2O} and \ce{N2}) as a mechanism that smooths and symmetrizes the nonlinear index modulation, which reduces the impact of spectral asymmetry and fine structure and enhances the fidelity of the compressed peak. The technique can be implemented with a capillary filled with ambient air, for sub-millijoule operation without a dedicated gas system. These results initiate a new direction in the optics of gas-filled waveguides and establish Raman-enhanced spectral compression as a robust route to high-energy narrowband optical sources, with potential impact in a broad range of applications.
\end{abstract}

\setboolean{displaycopyright}{false} 

\begin{document}
\maketitle

\section{Introduction}
Recent years have seen increasing interest in gas-based nonlinear optics, driven by its ability to access physical regimes that are difficult or impossible to realize in solid-state media \cite{Travers2011,Debord2019,Ferraro2025}. The choice of gas species provides a flexible platform for isolating and probing specific nonlinear processes, while the nonlinear response itself can be continuously tuned through the gas pressure. When implemented in capillaries or hollow-core fibers, the dispersion landscape becomes highly controllable as well, adjustable through both the gas pressure and the microstructure of the guiding geometry \cite{Marcatili1964}. Owing to the intrinsically-low nonlinear refractive index of gases, such systems are particularly well suited for high-energy frequency conversion \cite{Vicario2016,Carpeggiani2020,ZuritaMiranda2022} and for generating ultrashort pulses through nonlinear spectral broadening followed by temporal compression \cite{Nisoli1997,Travers2019,Grebing2020}. Research on gas-filled hollow-core waveguides has focused almost exclusively on spectral broadening and frequency-shifting.

Nonlinear spectral compression is a pulse-shaping technique in which the optical spectrum is intentionally narrowed through the interplay of self-phase modulation (SPM) and dispersion. This process enables efficient generation of narrowband high-brightness pulses and provides a complementary degree of freedom for tailoring ultrafast waveforms in nonlinear waveguides. It can help preserve the spectral integrity of individual channels in telecommunication fiber links \cite{Cundiff1999}, improve spectral resolution and signal-to-noise ratio in nonlinear Raman spectroscopy and microscopy \cite{Savvin2010}, or realize a coherent photonic interface between near-infrared quantum communication and near-visible quantum memory \cite{Li2017}. Spectral compression could also become a component of techniques for delivery of high-power short pulses through hollow-core fiber \cite{Clark2001}, which is an area of major current interest.

There are multiple routes to spectral compression. The earliest demonstrations relied on imposing a negative chirp on the input pulse before launching it into a nonlinear medium \cite{Stolen1978,Oberthaler1993,Planas1993}. During propagation, SPM generates a positive chirp that compensates for the negative pre-chirp, thereby flattening the frequency sweep and narrowing the optical spectrum. This has been demonstrated in optical solid-core fibers \cite{Oberthaler1993,Planas1993}, photonic crystal fibers \cite{Andresen2005}, glass plates \cite{Liu2006,Mitrofanov2018}, and silica-plate-based multipass cell \cite{Daher2020}. Other routes include adiabatic soliton evolution in an effectively dispersion-increasing fiber \cite{Chao2014,Nishizawa2010,Szewczyk2023}, soliton self-forming evolution under $0.5<N<1$ ($N$: soliton number) \cite{Sukiasyan2023}, self-similar evolution under exponentially-increasing dispersion \cite{Mei2024}, or sum frequency generation in a $\chi^{(2)}$ medium of two pulses exhibiting chirps of different signs \cite{Li2017}. All of these techniques are limited to pulse energies up to about \SI{10}{\micro\joule}.

Spectral compression could be valuable to the generation of high-energy transform-limited picosecond pulses, and gas-filled hollow-core fibers would be a natural platform for scaling the process to high energy. However, despite the significant effort devoted to study of pulse propagation in hollow-core fibers, there is only one report of spectral compression in them \cite{Lampen2022}, and its compressed spectrum is highly structured with a low compression ratio. Thus, an investigation of the process would also be a valuable contribution to the science of hollow-core fibers.

Here, we report the results of a theoretical and experimental investigation of spectral compression in a gas-filled capillary. Experiments were performed with molecular (\ce{N2O}, \ce{N2}) and noble (\ce{Ar}) gases to reveal the roles of rotational Raman and electronic nonlinearities in the process as well as the influence of the overall nonlinear response, which is enhanced in the molecular gases \cite{Beetar2020}. In contrast to prior works that focused only on the instantaneous electronic nonlinear response, we find that the delayed nature of the Raman nonlinear response in molecular gases \cite{Chen2024} plays a crucial and beneficial role in the compression process. Stimulated Raman scattering, occurring between the impulsive and transient Raman regimes, can induce an index change that smooths the temporal structure of a pulse. This enhances the quality of the compression, and enables \num{10} times compression with more than \SI{50}{\percent} of the pulse energy in the spectral peak in our experiments. In contrast, significant spectral pedestals are consistently observed for compression in \ce{Ar}, without the Raman nonlinearity. By varying the \ce{N2O} gas pressure, we demonstrate spectral compression for pulse energies ranging from \num{0.1} to \SI{1}{\milli\joule}. Furthermore, we applied this technique with ambient air as the nonlinear medium and achieved \num{9} times compression ratio for \SIadj{0.44}{\milli\joule} pulses. The air-based implementation significantly simplifies the setup while maintaining the performance, and holds promise for future applications that require high-energy, narrow-linewidth, and high-brightness laser sources.

\section{Raman-enhanced spectral compression}
\begin{figure}[!ht]
\centering
\includegraphics[width=\linewidth]{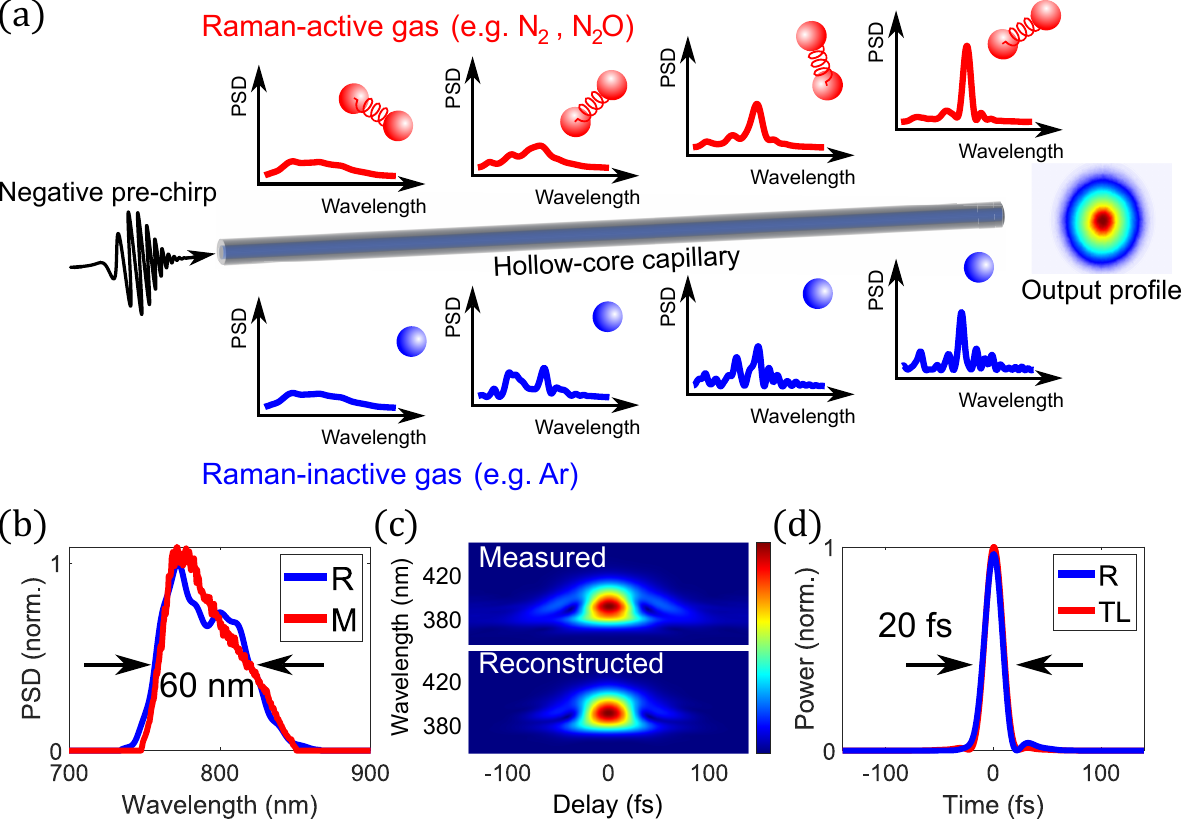}
\caption{(a) Illustration of the spectral-compression process in a capillary filled with Raman-active or -inactive gas. Insets show the spectral evolutions in Raman-active (red) or inactive (blue) gas, if the input spectrum has some structure. (b) Spectra of the input pulse independently measured with an optical spectral analyzer (M; red) and the retrieved pulse (R; blue) from our single-shot second-harmonic-generation frequency resolved optical gating (FROG) \cite{Mo2021}. The good agreement between them verifies the accuracy of the FROG measurement and retrieval. PSD: power spectral density. (c) Measured and reconstructed FROG traces of the input pulse. (d) Temporal profiles of the retrieved (R) and transform-limited (TL) input pulses.}
\label{fig:schematics}
\end{figure}

Fig.~\ref{fig:schematics}(a) depicts our experimental setup for the spectral compression. A \SIadj{20}{\fs} pulse [Figs.~\ref{fig:schematics}(b--d)] from a \SIadj{1}{\kilo\Hz} \ce{Ti}:sapphire laser system (Coherent Legend Elite HE+) was pre-chirped and launched into a \SIadj{1}{\m}-long gas-filled capillary. To avoid damage, we limited the launched pulse energy to at most \SI{1.4}{\milli\joule}. The chirped-pulse amplification design in the laser includes a Treacy-type grating dechirper, allowing us to precisely control the amount of applied negative pre-chirp to the pulse via fine adjustment of the grating separation. The capillary has a large core of \SIadj{400}{\micro\m} inner diameter to enable millijoule operations. The large core size also reduces the influence of dispersion \cite{Marcatili1964}, rendering an evolution dominated by nonlinearity for our investigations. The intensity profile of the capillary output displays a uniform near-Gaussian
distribution, which implies a dominant fundamental-mode evolution inside the capillary. The overall transmission of the capillary system is \SI{75}{\percent}, including the loss from the anti-reflection-coated input and uncoated output windows. The capillary was filled with Raman-active \ce{N2O} or \ce{N2}, or Raman-inactive \ce{Ar}. We select \ce{N2} due to the availability and safety considerations, and \ce{N2O} due to its (rotational) Raman transitions distributed around \SI{1}{\THz} \cite{Truong2024}, smaller than \SI{2}{\THz} in \ce{N2} \cite{Chen2024}, for a comparative study of the impact of the Raman response on the spectral-compression process.

For theoretical analysis, we performed numerical simulations based on the unidirectional pulse propagation equation with the realistic Raman response functions of Raman-active gases \cite{Chen2024}. Dispersion of the capillary is calculated by the Marcatili and Schmeltzer formula \cite{Marcatili1964}. Photoionization is implemented with the Perelomov-Popov-Terent'ev model \cite{Perelomov1966,Couairon2007} and we have verified that it does not have a significant impact on the spectral-compression process with the parameters investigated: although the \SIadj{1}{\milli\joule} energy of the unchirped pulse in our capillary would correspond to a small Keldysh parameter of about \num{1.3} in all three gases, the temporal stretching increases its value and thus mitigates peak-power-dependent photoionization. The simulations are conducted with the retrieved field [Fig.~\ref{fig:schematics}(d)] as the input, with nonlinear coefficient of gases adjusted to fit the experimental data as a calibration of our model.

\begin{figure}[!ht]
\centering
\includegraphics[width=\linewidth]{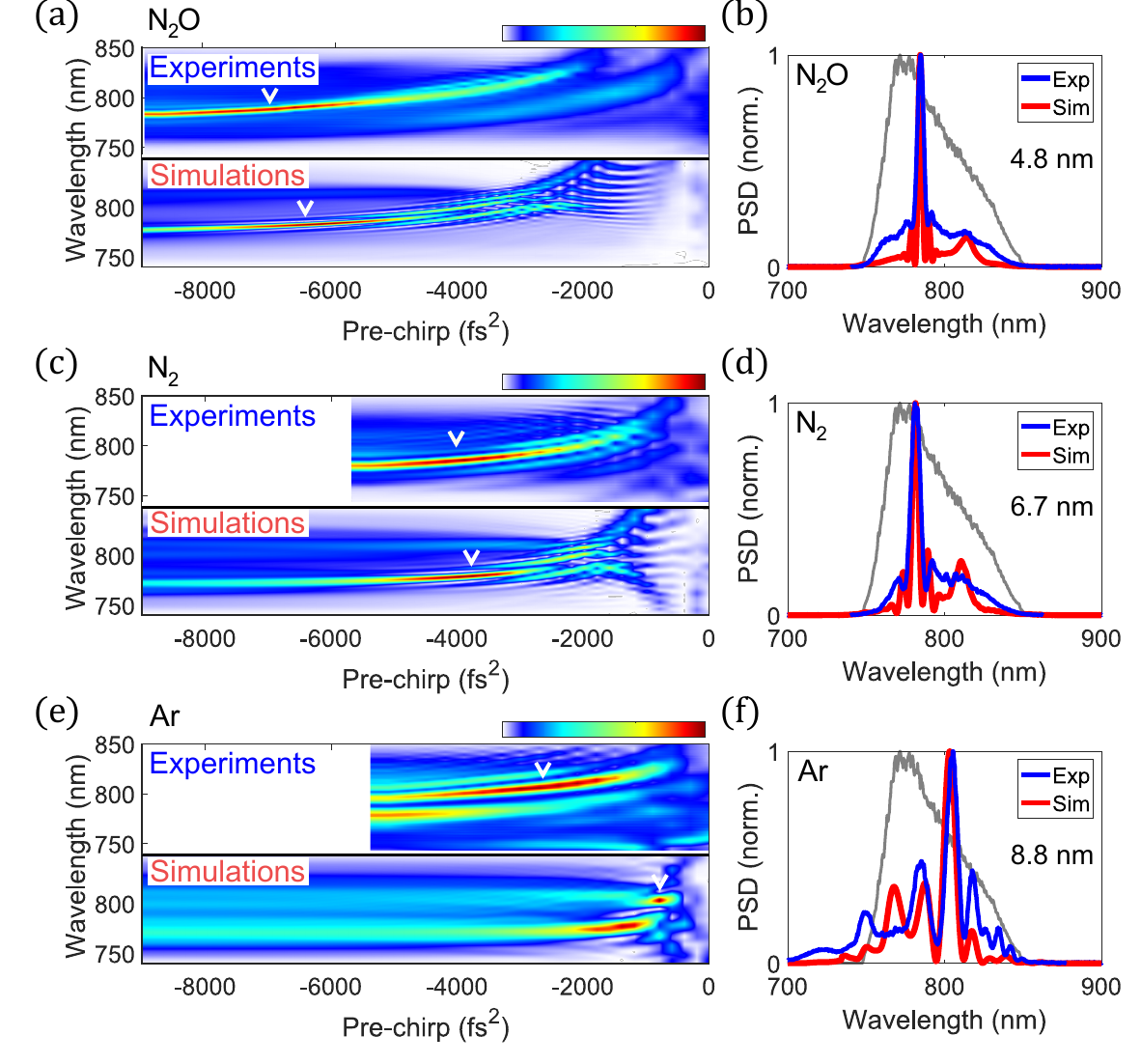}
\caption{(a) Measured and simulated spectra versus pre-chirp with \ce{N2O}. (b) shows the spectra from (a) that exhibits the strongest compression [labeled with tick symbols in (a)]. Gray line represents the input spectrum [red line in Fig.~\ref{fig:schematics}(b)]. The number on the right is the FWHM bandwidth of the compressed peak in the experiment. (c,d) and (e,f) are corresponding figures to (a,b) for \ce{N2} and \ce{Ar}, respectively. Injected pulse energies for \ce{N2O}, \ce{N2}, and \ce{Ar} are \num{0.13}, \num{0.48}, and \SI{0.42}{\milli\joule}, respectively. All three gas pressures are fixed at \SI{1}{\bar}. Measurements in \ce{N2} and \ce{Ar} are presented only up to around \SI{-5000}{\fs^2} due to the observed compression at smaller magnitude of pre-chirp. We found that the calibration factor of nonlinear coefficients for simulations to fit experiments is highly dependent on the FROG-retrieved fields we measured; for some measurements, the calibration factor is \num{1} for \ce{N2O} but can be up to \num{8} for \ce{Ar}. Here, calibrated nonlinear coefficients in all gases are twice as large as those in \cite{Chen2024,Bree2010} for the FROG-retrieved field in Fig.~\ref{fig:schematics}(b).}
\label{fig:first_demo}
\end{figure}

We first studied the spectral-compression process in the three gases at \SIadj{1}{\bar} pressure. Fig.~\ref{fig:first_demo} shows the variation of the output spectra for varying pre-chirp magnitude. In each case, the injected pulse energy was selected by optimizing the compression ratio. When the magnitude of the pre-chirp is small, the pulse experiences SPM, which broadens the spectrum. More negative pre-chirp leads to a narrowing spectrum. Beyond an optimal value of the pre-chirp, the spectrum broadens again. Unlike the structured spectra produced by \ce{Ar} [Figs.~\ref{fig:first_demo}(e,f)], propagation in \ce{N2O} and \ce{N2} generates a concentrated spectrum with smaller and smoother pedestals [Figs.~\ref{fig:first_demo}(a--d)]. The compression ratio increases with Raman strength, and reaches \num{9} in \ce{N2} and \num{12} in \ce{N2O}. The compressed peak is redshifted by \SIrange[range-phrase=--,range-units=single]{5}{10}{\nm} with respect to the peak of the input spectrum (at \SI{775}{\nm}) with \ce{N2O} and \ce{N2}, whereas it is redshifted by about \SI{30}{\nm} with \ce{Ar} (this will be discussed below). We also observe that the optimal compression occurs at a larger chirp magnitude in \ce{N2O} than in \ce{N2} [Figs.~\ref{fig:first_demo}(a,c)].

The experimental results exhibit clear spectral compression and also include some surprising features. Given that a down-chirped pulse and SPM are all that is required for spectral compression, one may wonder why stronger and cleaner compression is observed with the Raman-active gases. Numerical simulations that capture the main features of the measured spectra (Figs.~\ref{fig:first_demo}) provide some insights. The temporal dynamics of rotational Raman scattering play a major role in the experiments with \ce{N2O} and \ce{N2}. Due to the long dephasing time (\eg \SI{90}{\ps} for the rotational Raman transitions in \SIadj{1}{\bar} \ce{N2} \cite{Weber1994}), a sub-picosecond pulse undergoes impulsive or transient Raman scattering, depending on the pulse duration \cite{Chen2024}. When the pulse is shorter than the Raman transition period, Raman scattering is impulsively driven. Due to the rise time of the Raman response, impulsive Raman scattering produces an index change that is delayed relative to the intensity profile of the pulse. The delayed nonlinear response leads to a continuous redshift during propagation. If the pulse becomes longer than the Raman period, the Raman process enters the transient regime, where the Raman-induced index variation closely follows the driving pulse and thus contributes to SPM, without an overall redshift. In this regime, both the instantaneous electronic nonlinearity and the Raman nonlinearity contribute to SPM.

\begin{figure}[!ht]
\centering
\includegraphics[width=\linewidth]{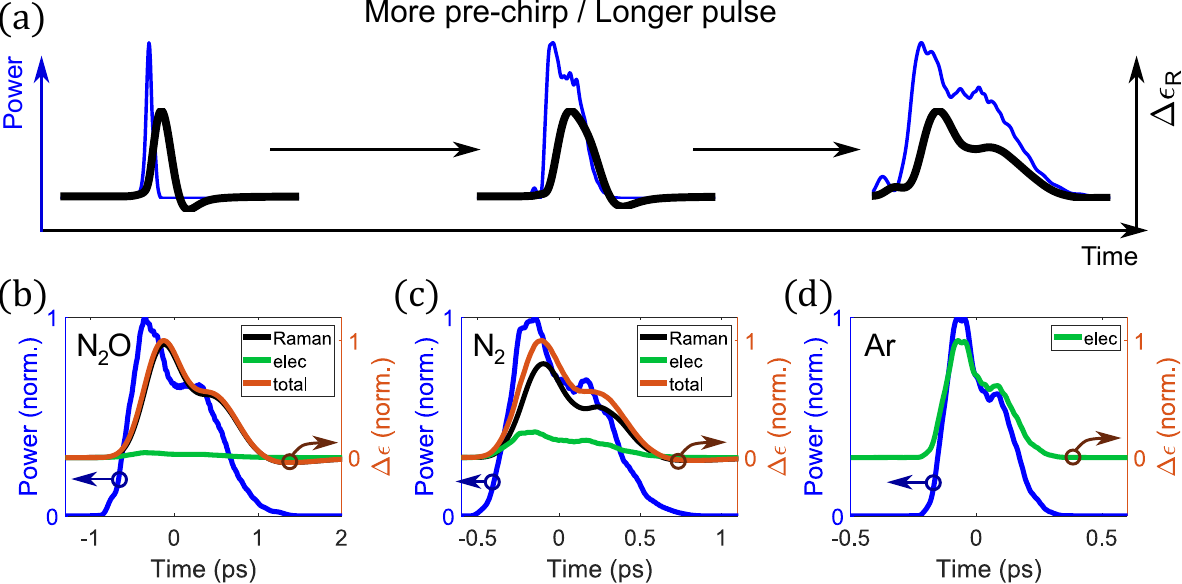}
\caption{(a) Raman-induced index change ($\triangle\epsilon_R$, black) for driving pulses (blue) with different stretched durations. (b--d) are simulated index changes induced by Raman (black) and electronic (green), as well as the total (orange), nonlinearities, in \ce{N2O}, \ce{N2}, and \ce{Ar}, respectively. Driving pulses (blue) are negatively chirped according to Figs.~\ref{fig:first_demo}(b,d,f).}
\label{fig:index_change}
\end{figure}

Spectral compression can be substantially enhanced by operating in the intermediate regime between the impulsive and transient limits, which is illustrated in the middle of Fig.~\ref{fig:index_change}(a). In this regime, the delayed Raman-induced index modulation roughly follows the temporal profile of the pulse, and reinforces the electronic SPM. At the same time, the slow Raman response acts as a temporal low-pass filter to produce an index modulation that is largely free of the fine temporal structure present in the driving pulse. In particular, if a pulse contains fine spectral features, temporal stretching through chirping maps these features onto the temporal profile \cite{Goda2009}. As the pulse duration increases, however, these temporal features are stretched accordingly, and the Raman-induced index modulation begins to replicate this extended structure. The resulting structured index profile disrupts the smooth nonlinear phase compensation required for efficient spectral compression \cite{Boscolo2018}. In addition to suppressing fine temporal structure, the Raman-induced index modulation in this intermediate regime partially decouples from the instantaneous pulse profile, reflecting instead the smoother Raman response function [see the left two figures in Fig.~\ref{fig:index_change}(a)]. This partial decoupling increases the symmetry of the resulting index modulation, and counters the asymmetry inherited from the driving pulse to some extent. The actual index changes for the three gases are shown in Figs.~\ref{fig:index_change}(b--d). These underlie the simulated spectra in Fig.~\ref{fig:first_demo}.

\begin{figure}[!ht]
\centering
\includegraphics[width=\linewidth]{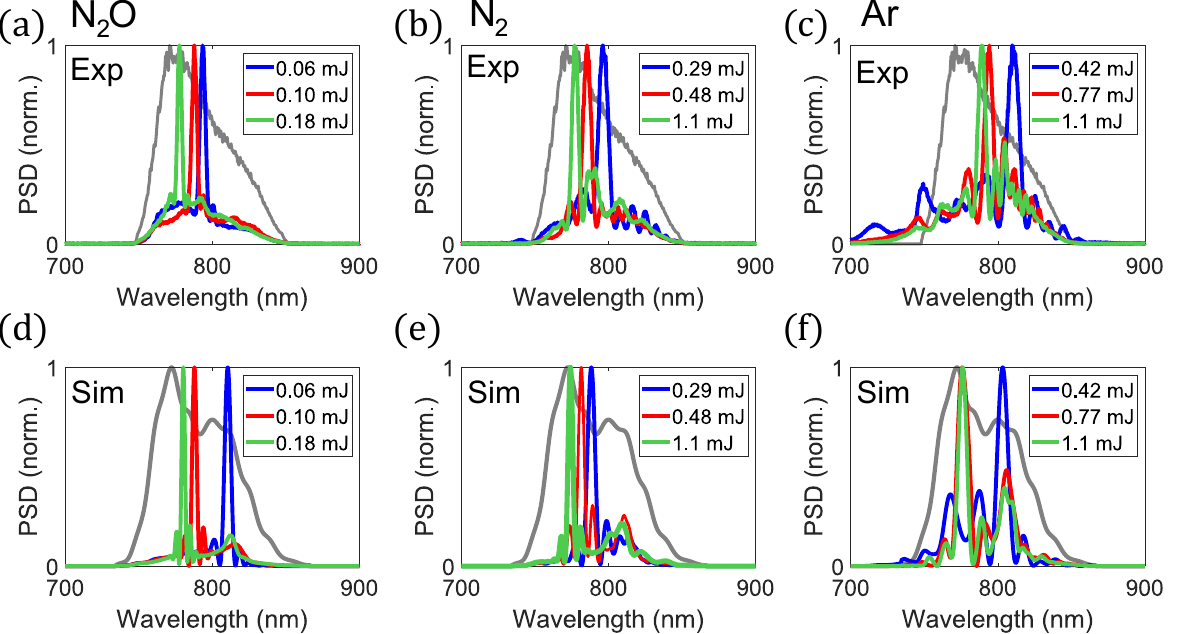}
\caption{(a--c) Experimental and (d--f) simulated output spectra of different injected energies in (left) \SIadj{1}{\bar} \ce{N2O}, (middle) \SIadj{1}{\bar} \ce{N2}, and (right) \SIadj{1}{\bar} \ce{Ar}. Gray lines in each figure represent the input spectra, showing the measured one in (a--c) and the retrieved one, considered in simulations, in (d--f). Pulses are negatively pre-chirped to obtain the optimal compression. Under a fixed gas pressure, the optimal chirp magnitude increases with pulse energy.}
\label{fig:peak_shift}
\end{figure}

Simulation results and the qualitative arguments above are consistent with the experimental results. Higher compression ratios and greater efficiency (defined as energy in the compressed peak divided by the total pulse energy) are observed for the molecular gases than for \ce{Ar}. The delayed Raman contribution, especially in \ce{N2O}, leads to a more-symmetric Gaussian-like (single-peak) index modulation within the pulse temporal extent [Figs.~\ref{fig:index_change}(b,c)]. On the contrary, the electronically-induced index modulation responds instantaneously to the driving pulse and therefore inherits all of its temporal structure. This makes spectral compression in \ce{Ar} susceptible to any temporal or spectral fine structure in the pulse and leads to degraded compressed results [as shown in Figs.~\ref{fig:first_demo}(f) and \ref{fig:index_change}(d)]. The apparent redshift of the compressed spectrum with \ce{Ar} is puzzling. The simulations for \ce{Ar} show that the output spectrum is influenced by the spectral structure of the driving pulse; relatively small changes in the input spectrum can produce peaks at wavelengths between \num{780} and \SI{810}{\nm} in the output spectrum. The secondary peak in the output corresponds to the \SIadj{810}{\nm} peak in the retrieved input spectrum [Fig.~\ref{fig:peak_shift}(f)], which also aligns with the positions of the compressed peaks in experiments [Fig.~\ref{fig:peak_shift}(c)]. We tentatively conclude that the underlying spectral structure of the input is the origin of the overall redshift observed in the compressed peak in \ce{Ar} in experiments [Fig.~\ref{fig:peak_shift}(c)]. The Raman response of \ce{N2O} is a factor of \num{2} slower than that of \ce{N2}, so we expect a similar increase in the stretched pulse duration to reach the intermediate Raman regime. This is observed: optimal compression occurs with a factor of \num{1.8} larger pre-chirp in \ce{N2O} than in \ce{N2} [Figs.~\ref{fig:first_demo}(a,c)]. In experiments with \ce{N2O} and \ce{N2}, the center wavelength of the compressed spectral peak depends on the input pulse duration. This is consistent with numerical simulations (Fig.~\ref{fig:peak_shift}) and can be understood as follows. With small chirp and short pulse duration, the delayed response extends across the pulse [left in Fig.~\ref{fig:index_change}(a)] and produces a large redshift. With increasing chirp and duration (\ie entering the transient Raman regime), the delay is small compared to the pulse duration, and thus the redshift is smaller. Moreover, due to the decoupling effect of the delayed Raman response, the peak of the Raman-induced index modulation is temporally shifted to the pulse center [Figs.~\ref{fig:index_change}(b,c)], compressing the spectrum around that center wavelength due to the pre-chirp-induced spectral-to-temporal mapping. It is worth noting that a similar trend of frequency shifting of the peak has been observed in experiments in \ce{Ar} [Fig.~\ref{fig:peak_shift}(c)]. The smooth input spectral profile in experiments causes the compressed peak to shift progressively toward this dominant component due to its strong nonlinear modulation after temporal mapping.

The Raman temporal response imposes an effective energy constraint with fixed gas pressure. As mentioned, higher-energy pulses require the introduction of larger pre-chirp to compensate for the increased nonlinear phase accumulated during propagation. A longer pulse can, in principle, yield a narrower spectral peak provided that its temporal phase is fully compensated. However, increasing the pulse duration also enhances the fine temporal structure in the nonlinear index modulation, which ultimately degrades the achievable spectral-compression performance. This degradation is evident in our search for the optimal compression conditions at different pulse energies in \ce{N2} [with a growing structure at \SI{1.1}{\milli\joule} in Fig.~\ref{fig:peak_shift}(b)]. Owing to the higher Raman fraction in \ce{N2O}, which suppresses index distortion, the degradation is reduced considerably [Fig.~\ref{fig:peak_shift}(a)].

\begin{figure}[!ht]
\centering
\includegraphics[width=\linewidth]{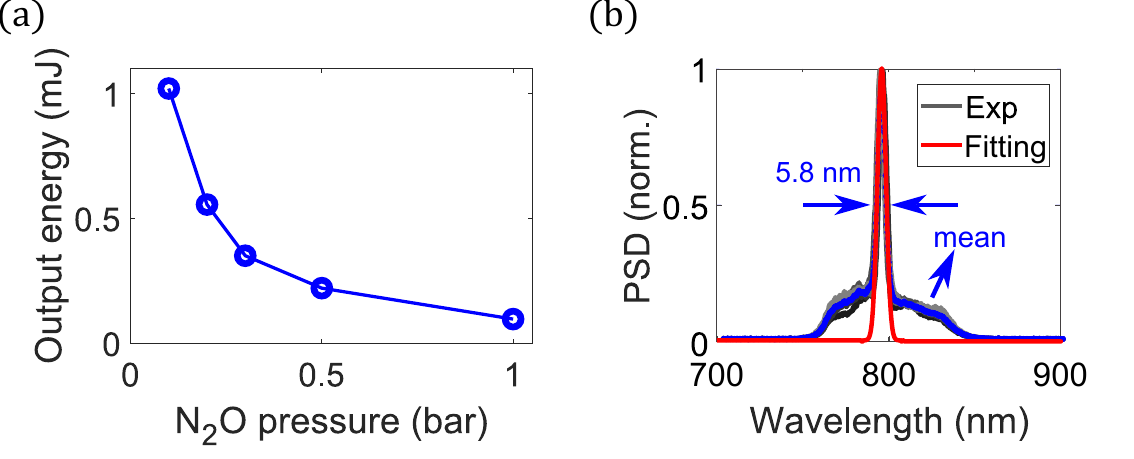}
\caption{(a) Variation of output energy at optimal spectral compression, under different \ce{N2O} gas pressures. (b) Spectra (five gray lines) after spectral compression in (a). A Gaussian fit (red) is applied to the mean spectrum (red) to estimate the energy fraction within the peak.}
\label{fig:high_energy_SC}
\end{figure}

To investigate the energy scaling of spectral compression, we conducted experiments at different gas pressures. A process with negligible dispersion is invariant under constant $\gamma\abs{A(t)}^2$, where $\gamma$ is the nonlinear coefficient (proportional to the gas pressure) and $A(t)$ is the input field. Because of the large core size of the capillary, dispersion remains negligible for the pressure range investigated, from \SI{-0.003}{\fs^2/\mm} under vacuum to \SI{0.017}{\fs^2/\mm} with \SI{1}{\bar} of \ce{N2O} \cite{Marcatili1964}. Thus, the process should only depend on the product of pressure and energy. By fixing the pre-chirp to \SI{-6000}{\fs^2}, which sets the minimum achievable spectral width through the corresponding stretched duration, we generated nearly-identical spectrally-compressed outputs across different \ce{N2O} pressures by maintaining a constant product of pressure and energy (Fig.~\ref{fig:high_energy_SC}). The spectral peak is \SIadj{5.8}{\nm} wide, for a compression ratio of \num{10}, and the energy fraction within is consistently around \SI{50}{\percent}. These results show that this process is scalable, and demonstrate scaling to \SI{1}{\milli\joule} with an injected energy of \SI{1.4}{\milli\joule}.

\section{Spectral compression in ambient air}
\begin{figure}[!ht]
\centering
\includegraphics[width=\linewidth]{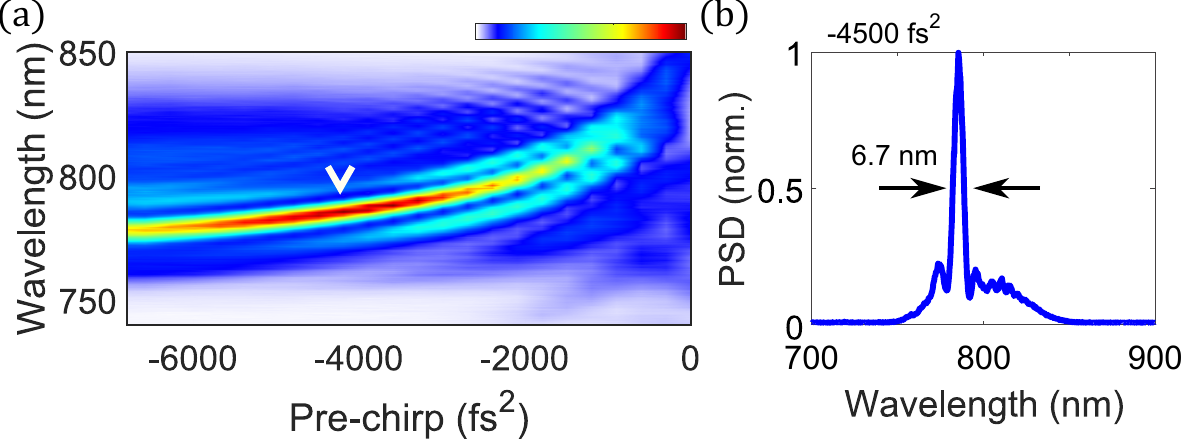}
\caption{(a) Measured spectra versus pre-chirp with ambient-air-filled capillary. (b) is the spectrum from (a) that exhibits the strongest spectral compression [labeled with a tick symbol in (a)].}
\label{fig:air}
\end{figure}

Although spectral compression can be realized in Raman-active gases such as \ce{N2O} and \ce{N2}, doing so necessitates a dedicated gas system, which introduces additional complexity to the overall experimental design. As air consists of \SI{78}{\percent} \ce{N2} and \SI{21}{\percent} \ce{O2}, both of which are Raman-active species, a significantly-simplified spectral-compression setup can be realized by using ambient air. Moreover, we expect that spectral compression in air performs similarly to that in pure \ce{N2} due to almost the same Raman line distributions of \ce{N2} and \ce{O2} \cite{Chen2024}. By launching \SIadj{0.44}{\milli\joule} pulses into an ambient-air-filled capillary, we generated a spectrally-compressed pulse with \SIadj{0.35}{\milli\joule} energy and \SIadj{6.7}{\nm} bandwidth, for a compression ratio of \num{9} (Fig.~\ref{fig:air}). This is the same performance as in \ce{N2} [Fig.~\ref{fig:first_demo}(d)]. Higher energy is possible with, however, increasing pedestal, also the same as that in pure \ce{N2} [Fig.~\ref{fig:peak_shift}(b)].

\section{Discussion}
A key outcome of this study is the demonstrated importance of maintaining a smooth and symmetric nonlinear index profile, so that the SPM-induced chirp is sufficiently linear to compensate for the applied linear negative pre-chirp. To date, investigations of spectral compression have primarily focused on the roles of dispersion \cite{Finot2016} and the initial temporal pulse shape \cite{Boscolo2018}. Imposing the pre-chirp on the pulse maps the spectral waveform into the time domain through a dispersive Fourier transform \cite{Goda2009}. Any fine structure present in the spectrum is thereby transferred to the temporal profile, where it perturbs the accumulated nonlinear phase and ultimately degrades the fidelity of the spectrally-compressed output. Prior studies employed solid waveguides, which generally benefit from the dispersion that suppresses the pedestal \cite{Finot2016}. On the other hand, for the plate-based multipass-cell implementation, which has low dispersion \cite{Daher2020}, no more than five-fold compression has been achieved before substantial pedestal growth becomes unavoidable. The single previous report of spectral compression in a gas employed the noble gas \ce{Kr} in a hollow-core fiber. That work generated a compressed spectrum more structured than the \ce{Ar} results we show here due to their structured input spectrum \cite{Lampen2022}. The experiments with \ce{Ar} in this work, even when driven by an input spectrum that is reasonably smooth [Fig.~\ref{fig:schematics}(b)], reveal the pronounced sensitivity of spectral compression to the fine structure in the pulse when the instantaneous electronic nonlinearity dominates.

\begin{figure}[!ht]
\centering
\includegraphics[width=\linewidth]{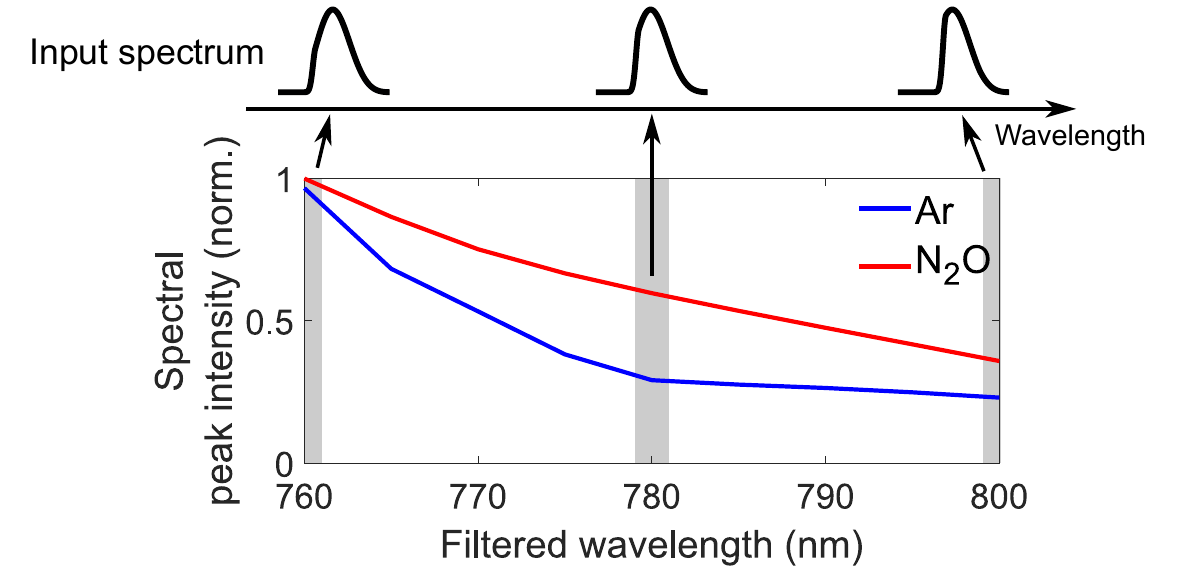}
\caption{Peak intensity of the simulated compressed spectra after propagating through \ce{Ar} or \ce{N2O}. A longpass filter is applied to the \SIadj{20}{\fs} pulse centered at \SI{800}{\nm} to introduce spectral asymmetry. A pre-chirp of \SI{-4000}{\fs^2} is imposed, smaller than the value used in Fig.~\ref{fig:first_demo}, in order to shorten the stretched duration and amplify the smoothing effect from the delayed Raman response. The fixed pre-chirp sets the Fourier-transform-limited spectral width, so enhanced spectral compression leads directly to higher spectral peak intensity.}
\label{fig:asymmetry}
\end{figure}

The use of Raman-active gases offers a naturally smoothed and somewhat-symmetrized index modulation that enhances the compressed peak fidelity. This smoothing mechanism emerges only when the interacting pulse drives Raman scattering between the impulsive and transient limits. Without Raman scattering, achieving comparable performance would require a laser source with a very-smooth symmetric bell-shaped, ideally parabolic, spectrum. Only a few prior works have examined the role of Raman scattering in spectral compression \cite{Liu2006,Fedotov2009}. These investigations focused on Raman-induced redshifting within a regime analogous to adiabatic soliton evolution, where spectral compression arises from increasing dispersion during the redshift in an optical fiber. In contrast, the present work identifies, for the first time, the decisive role of Raman temporal dynamics in shaping the overall nonlinear index profile that governs the quality of spectral compression. Several aspects of spectral compression in gases deserve further attention. Direct measurement of the temporal profile of the output pulse will be valuable; these were not possible in the present experiments owing to the \SIadj{500}{\fs} temporal window of our FROG system, which is designed for the measurement of very short pulses \cite{Mo2021}. The spectral-compression process in noble gases such as \ce{Ar} is certainly not well-understood, despite the nominal simplicity of only depending on the pre-chirp and electronic nonlinearity. We perform a simple analysis using a parabolic pulse as an input which is filtered to introduce spectral asymmetry (Fig.~\ref{fig:asymmetry}). The result shows enhanced spectral compression in \ce{N2O} compared to those in \ce{Ar}, but only under significant asymmetry. Further investigation is required to deliberately disrupt the spectral smoothness and quantify the influence of spectral irregularities. Finally, pedestals in the compressed spectra typically encompass more energy than predicted by simulations [Figs.~\ref{fig:peak_shift}(c,f)]. We speculate that this could arise from the presence of small satellite pulses or even amplified spontaneous emission in the input field, but more work is needed to resolve this issue.

Thermal deposition has been identified as a limiting factor for high-average-power operation in Raman-active gases, as elevated gas temperatures reduce the coherence-wave amplitude \cite{Beetar2021}. This limitation can be alleviated by using \ce{N2} instead of \ce{N2O}, since the faster Raman response of \ce{N2} produces a smaller overall redshift and approaches the transient regime \cite{Watson2025}, leading to fewer phonons generated, and thus less thermal deposition to the gas. Raman-based spectral compression demonstrated here generates less redshift as the pulse undergoes greater temporal stretching [Figs.~\ref{fig:peak_shift}(a,b,d,e)], which suggests that the impact of thermal deposition may be correspondingly reduced. Since additional temporal stretching can be achieved by increasing the gas pressure, there may exist an optimal operating point for each pulse energy at which Raman‑induced thermal effects are minimized. Future work will explore operation at repetition rates beyond the \SI{1}{\kilo\Hz} used in this study.

\section{Conclusion}
In conclusion, we have demonstrated a high-energy spectral-compression technique based on nonlinear propagation in gas-filled capillary, operating over (output) pulse energies from \num{0.1} to \SI{1}{\milli\joule}. The use of a noble gas for compression produces low compression ratios with structured spectra, and this is now understood to arise from mapping of spectral-to-temporal structure in the pre-chirping process. Raman-active gases offer stronger nonlinearities while smoothing the temporal nonlinear index change induced by the pulse. The method supports straightforward energy scaling through pressure tuning. Bandwidths below \SI{5}{\nm} and compression ratios above \num{10} are readily achieved, with approximately \SI{50}{\percent} of the energy in the compressed peak. Wavelength-tunability over a few spectral bandwidths is possible. Extension of this approach to ambient air enables spectral compression of sub-millijoule pulses without a dedicated gas-handling system. Additional energy scaling will be possible with larger-core capillaries. The demonstrated performance is a first step toward high-energy, narrowband optical sources for a range of scientific and technological applications. More broadly, the Raman physics identified here complements prior studies, in which the electronic nonlinearity dominates, and opens new avenues for gas-phase nonlinear optics research.

\begin{backmatter}
\bmsection{Funding} National Key Development Program for Basic Research of China (Grant Nos. 2024YFE0205800), the National Natural Science Foundation of China (Grant Nos. 12274158). U.S.\ National Institutes of Health (01EB033179, U01NS128660), the Office of U.S.\ Naval Research (N00014-19-1-2592).

\bmsection{Acknowledgment} Y.-H.C. was partially supported by a Mong Fellowship from Cornell Neurotech.

\bmsection{Disclosures} The authors declare no conflicts of interest.

\bmsection{Data availability} The code used in this work has been made publicly available at \url{https://github.com/AaHaHaa/gas_UPPE}.

\end{backmatter}

\bibliography{reference}

\begin{thebibliography}{10}
\newcommand{\enquote}[1]{``#1''}

\bibitem{Travers2011}
J.~C. Travers, W.~Chang, J.~Nold, \emph{et~al.}, \enquote{Ultrafast nonlinear optics in gas-filled hollow-core photonic crystal fibers {[Invited]},} {\protect\JournalTitle{J. Opt. Soc. Am. B}} \textbf{28}, A11--A26 (2011).

\bibitem{Debord2019}
B.~Debord, F.~Amrani, L.~Vincetti, \emph{et~al.}, \enquote{{Hollow-Core Fiber Technology: The Rising of “Gas Photonics”},} {\protect\JournalTitle{Fibers}} \textbf{7} (2019).

\bibitem{Ferraro2025}
M.~Ferraro, B.~Kibler, P.~B\'{e}jot, \emph{et~al.}, \enquote{Extreme nonlinear optics in optical fibers,} {\protect\JournalTitle{arXiv preprint arXiv:2512.25046}}  (2025).

\bibitem{Marcatili1964}
E.~A.~J. Marcatili and R.~A. Schmeltzer, \enquote{Hollow metallic and dielectric waveguides for long distance optical transmission and lasers,} {\protect\JournalTitle{Bell Syst. Tech. J.}} \textbf{43}, 1783--1809 (1964).

\bibitem{Vicario2016}
C.~Vicario, M.~Shalaby, A.~V. Konyashchenko, \emph{et~al.}, \enquote{High-power femtosecond {Raman} frequency shifter,} {\protect\JournalTitle{Opt. Lett.}} \textbf{41}, 4719--4722 (2016).

\bibitem{Carpeggiani2020}
P.~A. Carpeggiani, G.~Coccia, G.~Fan, \emph{et~al.}, \enquote{Extreme {Raman} red shift: ultrafast multimode nonlinear space-time dynamics, pulse compression, and broadly tunable frequency conversion,} {\protect\JournalTitle{Optica}} \textbf{7}, 1349--1354 (2020).

\bibitem{ZuritaMiranda2022}
O.~Zurita-Miranda, C.~Fourcade-Dutin, F.~Fauquet, \emph{et~al.}, \enquote{Tunable ultrafast infrared generation in a gas-filled hollow-core capillary by a four-wave mixing process,} {\protect\JournalTitle{J. Opt. Soc. Am. B}} \textbf{39}, 662--670 (2022).

\bibitem{Nisoli1997}
M.~Nisoli, S.~D. Silvestri, O.~Svelto, \emph{et~al.}, \enquote{Compression of high-energy laser pulses below {5~fs},} {\protect\JournalTitle{Opt. Lett.}} \textbf{22}, 522--524 (1997).

\bibitem{Travers2019}
J.~C. Travers, T.~F. Grigorova, C.~Brahms, and F.~Belli, \enquote{High-energy pulse self-compression and ultraviolet generation through soliton dynamics in hollow capillary fibres,} {\protect\JournalTitle{Nat. Photon.}} \textbf{13}, 547--554 (2019).

\bibitem{Grebing2020}
C.~Grebing, M.~M\"{u}ller, J.~Buldt, \emph{et~al.}, \enquote{Kilowatt-average-power compression of millijoule pulses in a gas-filled multi-pass cell,} {\protect\JournalTitle{Opt. Lett.}} \textbf{45}, 6250--6253 (2020).

\bibitem{Cundiff1999}
S.~T. Cundiff, B.~C. Collings, L.~Boivin, \emph{et~al.}, \enquote{{Propagation of Highly Chirped Pulses in Fiber-Optic Communications Systems},} {\protect\JournalTitle{J. Lightwave Technol.}} \textbf{17}, 811 (1999).

\bibitem{Savvin2010}
A.~D. Savvin, A.~A. Lanin, A.~A. Voronin, \emph{et~al.}, \enquote{Coherent {anti-Stokes} {Raman} metrology of phonons powered by photonic-crystal fibers,} {\protect\JournalTitle{Opt. Lett.}} \textbf{35}, 919--921 (2010).

\bibitem{Li2017}
Y.~Li, T.~Xiang, Y.~Nie, \emph{et~al.}, \enquote{Spectral compression of single-photon-level laser pulse,} {\protect\JournalTitle{Sci. Rep.}} \textbf{7}, 43494 (2017).

\bibitem{Clark2001}
S.~W. Clark, F.~O. Ilday, and F.~W. Wise, \enquote{Fiber delivery of femtosecond pulses from a {$\mathrm{Ti}$}:sapphire laser,} {\protect\JournalTitle{Opt. Lett.}} \textbf{26}, 1320--1322 (2001).

\bibitem{Stolen1978}
R.~H. Stolen and C.~Lin, \enquote{Self-phase-modulation in silica optical fibers,} {\protect\JournalTitle{Phys. Rev. A}} \textbf{17}, 1448--1453 (1978).

\bibitem{Oberthaler1993}
M.~Oberthaler and R.~A. H\"{o}pfel, \enquote{Special narrowing of ultrashort laser pulses by self‐phase modulation in optical fibers,} {\protect\JournalTitle{Appl. Phys. Lett.}} \textbf{63}, 1017--1019 (1993).

\bibitem{Planas1993}
S.~A. Planas, N.~L.~P. Mansur, C.~H.~B. Cruz, and H.~L. Fragnito, \enquote{Spectral narrowing in the propagation of chirped pulses in single-mode fibers,} {\protect\JournalTitle{Opt. Lett.}} \textbf{18}, 699--701 (1993).

\bibitem{Andresen2005}
E.~R. Andresen, J.~Th{\o}gersen, and S.~R. Keiding, \enquote{Spectral compression of femtosecond pulses in photonic crystal fibers,} {\protect\JournalTitle{Opt. Lett.}} \textbf{30}, 2025--2027 (2005).

\bibitem{Liu2006}
J.~Liu, X.~Chen, J.~Liu, \emph{et~al.}, \enquote{Spectrum reshaping and pulse self-compression in normally dispersive media with negatively chirped femtosecond pulses,} {\protect\JournalTitle{Opt. Express}} \textbf{14}, 979--987 (2006).

\bibitem{Mitrofanov2018}
A.~V. Mitrofanov, M.~M. Nazarov, A.~A. Voronin, \emph{et~al.}, \enquote{Free-beam spectral self-compression at supercritical peak powers,} {\protect\JournalTitle{Opt. Lett.}} \textbf{43}, 5693--5696 (2018).

\bibitem{Daher2020}
N.~Daher, F.~Guichard, X.~D\'{e}len, \emph{et~al.}, \enquote{Spectral compression in a multipass cell,} {\protect\JournalTitle{Opt. Express}} \textbf{28}, 21571--21577 (2020).

\bibitem{Chao2014}
W.-T. Chao, Y.-Y. Lin, J.-L. Peng, and C.-B. Huang, \enquote{Adiabatic pulse propagation in a dispersion-increasing fiber for spectral compression exceeding the fiber dispersion ratio limitation,} {\protect\JournalTitle{Opt. Lett.}} \textbf{39}, 853--856 (2014).

\bibitem{Nishizawa2010}
N.~Nishizawa, K.~Takahashi, Y.~Ozeki, and K.~Itoh, \enquote{Wideband spectral compression of wavelength-tunable ultrashort soliton pulse using comb-profile fiber,} {\protect\JournalTitle{Opt. Express}} \textbf{18}, 11700--11706 (2010).

\bibitem{Szewczyk2023}
O.~Szewczyk, Z.~Łaszczych, and G.~Sobo\'{n}, \enquote{Spectral compression and amplification of ultrashort pulses tunable in the 1650 – 1900~nm wavelength range,} {\protect\JournalTitle{Opt. Laser Technol.}} \textbf{164}, 109465 (2023).

\bibitem{Sukiasyan2023}
M.~Sukiasyan, V.~Avetisyan, and A.~Kutuzyan, \enquote{{Spectral Self-Compression of Chirp-Free Pulses in Anomalously Dispersive Optical Fibers},} {\protect\JournalTitle{Photonics}} \textbf{10} (2023).

\bibitem{Mei2024}
C.~Mei, Y.~Zhang, X.~Zhou, and H.-G. Duan, \enquote{On the theory of spectral compression-assisted optical temporal differentiation,} {\protect\JournalTitle{Opt. Express}} \textbf{32}, 43146--43160 (2024).

\bibitem{Lampen2022}
J.~Lampen, F.~Tani, P.~Li, \emph{et~al.}, \enquote{{Spectral Self-Compression in Gas-Filled Hollow-Core Photonic Crystal Fiber},} in \emph{Conference on Lasers and Electro-Optics,}  (Optica Publishing Group, 2022), p. SW4K.3.

\bibitem{Beetar2020}
J.~E. Beetar, M.~Nrisimhamurty, T.-C. Truong, \emph{et~al.}, \enquote{Multioctave supercontinuum generation and frequency conversion based on rotational nonlinearity,} {\protect\JournalTitle{Sci. Adv.}} \textbf{6}, eabb5375 (2020).

\bibitem{Chen2024}
Y.-H. Chen and F.~Wise, \enquote{Unified and vector theory of {Raman} scattering in gas-filled hollow-core fiber across temporal regimes,} {\protect\JournalTitle{APL Photon.}} \textbf{9}, 030902 (2024).

\bibitem{Truong2024}
T.-C. Truong, C.~Lantigua, Y.~Zhang, \emph{et~al.}, \enquote{{Spectral Broadening and Pulse Compression in Molecular Gas-Filled Hollow-Core Fibers},} {\protect\JournalTitle{IEEE J. Sel. Topics Quantum Electron.}} \textbf{30}, 1--11 (2024).

\bibitem{Mo2021}
Y.~Mo, W.~Cao, H.~Xu, \emph{et~al.}, \enquote{Few-cycle optical pulse characterization under phase-mismatching,} {\protect\JournalTitle{Opt. Lett.}} \textbf{46}, 548--551 (2021).

\bibitem{Perelomov1966}
A.~M. {Perelomov}, V.~S. {Popov}, and M.~V. {Terent'ev}, \enquote{{Ionization of Atoms in an Alternating Electric Field},} {\protect\JournalTitle{Sov. Phys. JETP}} \textbf{23}, 924 (1966).

\bibitem{Couairon2007}
A.~Couairon and A.~Mysyrowicz, \enquote{Femtosecond filamentation in transparent media,} {\protect\JournalTitle{Phys. Rep.}} \textbf{441}, 47--189 (2007).

\bibitem{Bree2010}
C.~Bree, A.~Demircan, and G.~Steinmeyer, \enquote{{Method for Computing the Nonlinear Refractive Index via Keldysh Theory},} {\protect\JournalTitle{IEEE J. Quantum Electron.}} \textbf{46}, 433--437 (2010).

\bibitem{Weber1994}
M.~J. Weber, \emph{{CRC Handbook of Laser Science and Technology Supplement 2}} (CRC Press, 1994).

\bibitem{Goda2009}
K.~Goda, D.~R. Solli, K.~K. Tsia, and B.~Jalali, \enquote{Theory of amplified dispersive {Fourier} transformation,} {\protect\JournalTitle{Phys. Rev. A}} \textbf{80}, 043821 (2009).

\bibitem{Boscolo2018}
S.~Boscolo, F.~Chaussard, E.~Andresen, \emph{et~al.}, \enquote{Impact of initial pulse shape on the nonlinear spectral compression in optical fibre,} {\protect\JournalTitle{Opt. Laser Technol.}} \textbf{99}, 301--309 (2018).

\bibitem{Finot2016}
C.~Finot and S.~Boscolo, \enquote{Design rules for nonlinear spectral compression in optical fibers,} {\protect\JournalTitle{J. Opt. Soc. Am. B}} \textbf{33}, 760--767 (2016).

\bibitem{Fedotov2009}
A.~B. Fedotov, A.~A. Voronin, I.~V. Fedotov, \emph{et~al.}, \enquote{Spectral compression of frequency-shifting solitons in a photonic-crystal fiber,} {\protect\JournalTitle{Opt. Lett.}} \textbf{34}, 662--664 (2009).

\bibitem{Beetar2021}
J.~E. Beetar, M.~Nrisimhamurty, T.-C. Truong, \emph{et~al.}, \enquote{Thermal effects in molecular gas-filled hollow-core fibers,} {\protect\JournalTitle{Opt. Lett.}} \textbf{46}, 2437--2440 (2021).

\bibitem{Watson2025}
K.~Watson, T.~Saule, M.~Ivanov, \emph{et~al.}, \enquote{High-power femtosecond molecular broadening and the effects of ro-vibrational coupling,} {\protect\JournalTitle{Optica}} \textbf{12}, 5--10 (2025).

\end{thebibliography}

\bibliographyfullrefs{reference}

\end{document}